\documentclass[letter,letterpage]{IEEEtran}
\IEEEoverridecommandlockouts
\usepackage{cite}
\usepackage{amsmath,amssymb,amsfonts}
\usepackage{algorithmic}
\usepackage{graphicx}
\usepackage{multirow}
\usepackage{textcomp}
\usepackage{xcolor}
\usepackage{glossaries}
\usepackage{textcomp}
\usepackage{svg}
\usepackage{booktabs}
\usepackage[inline]{enumitem}
\usepackage{siunitx}
\usepackage{overpic}
\usepackage[symbol]{footmisc}

\usepackage[hidelinks,colorlinks=false]{hyperref}%
\usepackage[capitalise]{cleveref}

\newcommand{\ra}[1]{\renewcommand{\arraystretch}{#1}}
\def\BibTeX{{\rm B\kern-.05em{\sc i\kern-.025em b}\kern-.08em
    T\kern-.1667em\lower.7ex\hbox{E}\kern-.125emX}}

\newacronym{dvs}{DVS}{dynamic vision sensor}
\newacronym{erc}{ERC}{Event Rate Control}
\newacronym{fov}{FoV}{field of view}
\newacronym{rgb}{RGB}{color}
\newacronym{ai}{AI}{artificial intelligence}
\newacronym{dnn}{DNN}{deep neural network}
\newacronym{nn}{NN}{neural network}
\newacronym{cots}{COTS}{commercial off the shelf}
\newacronym{snn}{SNN}{spiking neural network}
\newacronym{sota}{SOTA}{state-of-the-art}
\newacronym{cmos}{CMOS}{complementary metal-oxide semiconductor}
\newacronym{ccd}{CCD}{charge-coupled device}
\newacronym{zncc}{ZNCC}{zero mean normalized cross-correlation}

\usepackage{eso-pic}
\usepackage{url}
\AddToShipoutPictureBG*{
  \AtPageUpperLeft{%
    \put(0,-40){\raisebox{15pt}{\makebox[\paperwidth]{\begin{minipage}{21cm}\centering
      \textcolor{gray}{This paper has been accepted for publication at \\ IEEE SENSORS, Vienna, 2023.}
    \end{minipage}}}}%
  }
  \AtPageLowerLeft{%
    \raisebox{20pt}{\makebox[\paperwidth]{\begin{minipage}{21cm}\centering
      \textcolor{gray}{ \copyright 2023 IEEE.  Personal use of this material is permitted.  Permission from IEEE must be obtained for all other uses, in any current or future media, including reprinting/republishing this material for advertising or promotional purposes, creating new collective works, for resale or redistribution to servers or lists, or reuse of any copyrighted component of this work in other works.
      }
    \end{minipage}}}%
  }
}

\begin{document}
\bstctlcite{IEEEexample:BSTcontrol} 

\newcolumntype{M}[1]{>{\centering\arraybackslash}m{#1}}

\title{Quantitative Evaluation of a Multi-Modal Camera Setup for Fusing Event Data with RGB Images}  

\author{
\IEEEauthorblockN{
Julian Moosmann, 
Jakub Mandula, 
Philipp Mayer, 
Luca Benini, 
Michele Magno} 
\IEEEauthorblockA{\\Dept. of Information Technology and Electrical Engineering, ETH Z\"{u}rich, Switzerland} 
}


\maketitle
\thispagestyle{empty}
\begin{abstract}
Event-based cameras, also called silicon retinas, potentially revolutionize computer vision by detecting and reporting significant changes in intensity asynchronous events, offering extended dynamic range, low latency, and low power consumption, enabling a wide range of applications from autonomous driving to longtime surveillance. As an emerging technology, there is a notable scarcity of publicly available datasets for event-based systems that also feature frame-based cameras, in order to exploit the benefits of both technologies. 
This work quantitatively evaluates a multi-modal camera setup for fusing high-resolution \gls{dvs} data with RGB image data by static camera alignment. The proposed setup, which is intended for semi-automatic \gls{dvs} data labeling, combines two recently released Prophesee EVK4 \gls{dvs} cameras and one global shutter XIMEA MQ022CG-CM RGB camera. After alignment, state-of-the-art object detection or segmentation networks label the image data by mapping boundary boxes or labeled pixels directly to the aligned events. To facilitate this process, various time-based synchronization methods for DVS data are analyzed, and calibration accuracy, camera alignment, and lens impact are evaluated.
Experimental results demonstrate the benefits of the proposed system: the best synchronization method yields an image calibration error of less than \SI{0.90}{px} and a pixel cross-correlation deviation of \SI{1.6}{px}, while a lens with \SI{8}{\mm} focal length enables detection of objects with size \SI{30}{\centi\m} at a distance of \SI{350}{\m} against homogeneous background.
\end{abstract}
\glsreset{dvs}


%
\section{Introduction}
\begin{figure}
    \centering
    \begin{overpic}[width=\columnwidth]{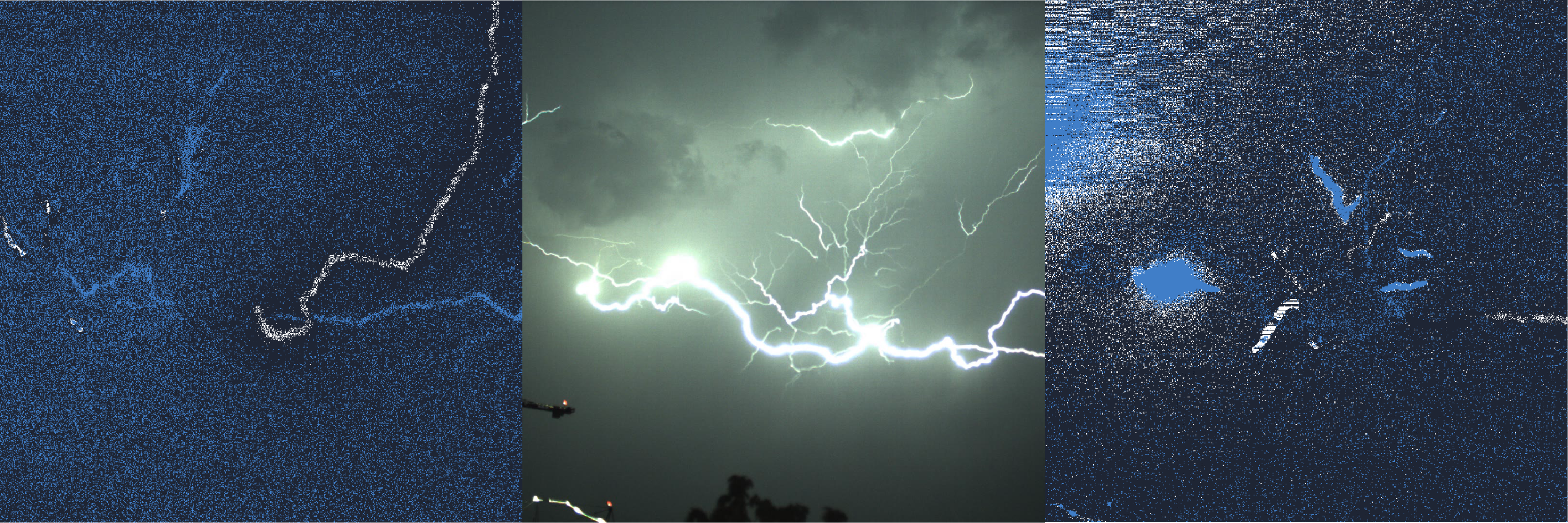}
        \put(28,29.5){\textcolor{white}{(a)}}
        \put(61,29.5){\textcolor{white}{(b)}}
        \put(94.5,29.5){\textcolor{white}{(c)}}
    \end{overpic}
    \caption{\Gls{dvs} camera recording a bolt of lightning (a) clear image without saturation. (b) The corresponding RGB image. (c) Shows the sensor saturating and merging events together into spurious line artifacts.}
    \vspace{-2em}
    \label{fig:lightning}
\end{figure}

Deep learning and \gls{ai} have accelerated and changed the way computers process visual information in digital systems \cite{chai2021deep}. Image classification \cite{chen2021review}, object detection \cite{carion2020eccv, liu2016eccv}, or semantic segmentation \cite{long2015cvpr} are essential capabilities for autonomous agents with tasks such as localization \cite{zimmerman2022long}, mapping \cite{rosinol2020kimera, zimmerman2023long}, or navigation \cite{crespo2020semantic}.
Most algorithms for object detection and semantic segmentation operate on \gls{rgb} image frames\cite{liu2017survey}. Notably, these algorithms heavily rely on extensive data for training the underlying \gls{nn} \cite{zhu2016we}. Unfortunately, the current commercially available \gls{ccd} and \gls{cmos} sensors struggle with high dynamic range and produce a vast quantity of dense data making the tracking of high-velocity objects in adverse lighting conditions a challenge \cite{el2005cmos, hillebrand2000high, kao2006integrating}. To address this limitation, event-based cameras, also known as silicon retinas, have emerged as a promising bio-inspired branch of image sensors \cite{gallego2020event, zheng2023deep}. \Gls{dvs} cameras do not directly measure the absolute intensity of light but rather detect and respond to changes in light intensity as seen in \cref{fig:lightning}. When a DVS pixel detects an intensity change, the camera generates an asynchronous message with the pixel coordinates. This differs from the conventional approach of utilizing a rolling or global shutter mechanism and allows for an exceptional dynamic range \cite{lichtsteiner2008128} of over \SI{86}{\decibel} \cite{EventbasedVisionSensor}  and superior capturing speed \cite{gallego2020event}. As such the \gls{dvs} cameras produce highly sparse data, especially in static scenarios.

However, the novelty of the technology results in a scarcity of available labeled datasets containing \gls{dvs} data \cite{lakshmi2019neuromorphic}, and even fewer datasets are available with a dual camera scenario where \gls{dvs} are combined with RGB cameras. The necessity to create new algorithms that account for time dependencies is approached by two main ideas. A bio-inspired data processing approach using \glspl{snn} which requires specialized hardware from the field of mixed analog and digital processing \cite{2023negiBestBothWorlds}. Or a frame-based approach that allows for the utilization of traditional \glspl{dnn} by accumulating events and integrating them over a certain time period and as such creates a two-dimensional image frame out of event data \cite{2018maquedaEventbasedVisionMeets}. This work presents and evaluates a system for \gls{dvs} dataset collection based on simultaneous recordings of RGB frames and \gls{dvs} data. Subsequently, postprocessing techniques are utilized to align the RGB and \gls{dvs} pixels. Once aligned, a state-of-the-art object detection algorithm such as YOLO \cite{redmon2016you, redmon2017yolo9000, redmon2018yolov3, bochkovskiy2020yolov4, Jocher_YOLOv5_by_Ultralytics_2020, li2022yolov6, wang2023yolov7, Jocher_YOLO_by_Ultralytics_2023} is applied to the RGB frames, yielding accurate pixel/box labels that can be transformed onto the \gls{dvs} camera view as proposed in \cite{perot2020learning}. \\ The main contributions of the work can be summarized as follows:
\begin{enumerate*}[label=(\roman*),,font=\itshape]
    \item Investigation of data synchronization methods for frame- and event-based vision sensors, achieving a calibration accuracy with a standard deviation of \SI{0.90}{px} and projected image cross-correlation deviation of \SI{1.6}{px}.
    \item Quantitative analysis of object perception by combining \gls{dvs} sensors with lenses of different focal lengths.
    \item Practical evaluation of data generation quantity for various vision scenarios for a 1280$\times$720 pixels \gls{dvs} sensor.
\end{enumerate*}

\begin{figure}[t]
\centerline{\includegraphics[width=0.7\columnwidth]{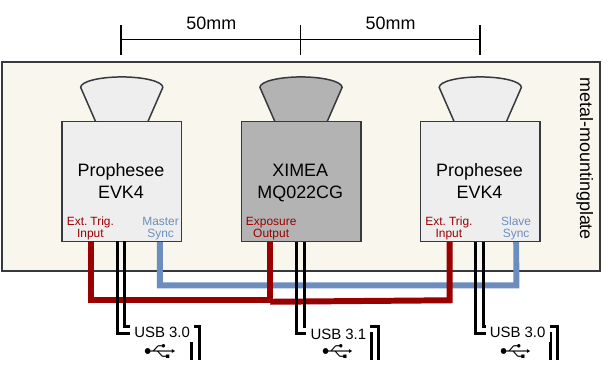}}
\caption{The proposed hardware setup for data collection includes two Prophesee EVK4 event-based cameras and one Ximea RGB image sensor (MQ022CG-CM) with hardware synchronization. The Ximea outputs a digital signal during its exposure time that is hardware time-stamped into the event stream.}
\label{fig:hw_setup}
    \vspace{-1em}
\end{figure}

\begin{table}[b!]
    \vspace{-1em}
    \ra{1.1}
    \caption{An overview of the utilized C-mount lenses.}
     \centering
    \resizebox{\linewidth}{!}{%
    \begin{tabular}{@{}lllrr@{}}\toprule
        & & \multicolumn{3}{c}{\textbf{Specs}}\\
        \cmidrule{2-5}
        \textbf{Camera} & \multicolumn{2}{c}{\textbf{Lens}} & \textbf{Focal lenght}$^*$  & \textbf{Distortion}\\
        \midrule
        \multirow{4}{*}{\rotatebox[origin=c]{90}{\textbf{Ximea}}} & X100 & Navitar: MVL100M23 & \SI{100}{mm} & \SI{0.05}{\percent}\\      
        & X75 & Tamron: M112FM75 & \SI{75}{mm} & \SI{0.00}{\percent}\\      
        & X50 & Computar: M5028-MPW3 & \SI{50}{mm} & \SI{0.00}{\percent}\\
        & X12.5 & Fujinon: HF12.5SA-1 & \SI{12.5}{mm} & \SI{0.40}{\percent}\\
        \midrule
        \multirow{4}{*}{\rotatebox[origin=c]{90}{\textbf{Prophesee}}} & P100 & Navitar: MVL100M23 & \SI{100(181)}{mm} & \SI{0.05}{\percent}\\
        & P75 & Computar: M7528-MPW3 & \SI{75(136)}{mm} & \SI{0.12}{\percent}\\    
        & P50 & Tamron: M117FM50-RG & \SI{50(91)}{mm} & \SI{0.01}{\percent}\\      
        & P35 & Computar: M3528-MPW3 & \SI{35(63)}{mm} & \SI{0.03}{\percent}\\ 
        & P8 & SOYO: SFA0820-5M & \SI{8(15)}{mm} & \SI{<0.1}{\percent}\\ 
        \bottomrule
        \multicolumn{5}{l}{$^*$ Number in (brackets) is the effective focal length accounting for the sensor's crop factor.} 
    \end{tabular}
    \label{tab:lens_overview}
    }
\end{table}
\vspace{-0.25cm}
\section{Camera Setup}\label{sec:setup}
The proposed multi-camera setup (\cref{fig:hw_setup}) consists of one frame-based \gls{cmos} \gls{rgb} camera from Ximea (MQ022CG-CM) and two Prophesee EVK4 event cameras---having the largest \gls{dvs} resolution commercially available---all of which are hardware synchronized to each other.
The two Prophesee cameras enable stereo vision capture, however, in this work, different optics were mounted on each camera in order to efficiently compare lenses with various \glspl{fov}. The primary function of the \gls{rgb} camera is to facilitate semi-automatic labeling, which aids in the creation of event-based datasets. Calibration is utilized to align the various camera views and allow labels to be transformed between them as suggested in \cite{de2020large}. To simplify the subsequent alignment with event data, a global shutter RGB camera has been chosen to mitigate the complications arising from the row-wise reading of pixels in the rolling shutter cameras. In addition to the targeted field of view and the varying-sized camera sensors\footnote{Prophesee: EVK4 has a sensor size of 1/2.5" and a resolution of 1280$\times$720 while the Ximea: MQ022CG-CM global shutter \gls{cmos} camera has a sensor size of 2/3" and a resolution of 2048$\times$1088}, the lens' focal lengths had to be carefully chosen in order to closely align the \gls{fov} and minimize the distortion. In particular, the lenses were chosen such that the Ximea camera has slightly larger \gls{fov} allowing for full coverage of the event camera \gls{fov}. \Cref{tab:lens_overview} outlines the final utilized lenses and \cref{tab:lens_combinations} the lens combinations.

\begin{figure}[t]
    \centering
    \includegraphics[width=0.8\columnwidth]{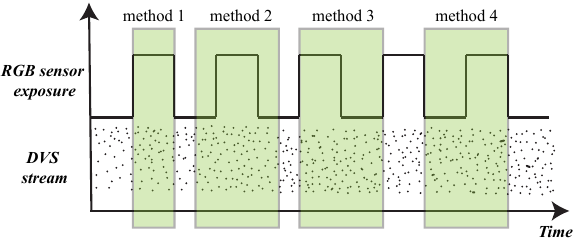}
    \caption{Visualization of the four ways to synchronize event- with frame-based data. The green windows illustrate which events are integrated and synchronized to an RGB image frame. An RGB image frame is created whenever the RGB sensor exposes, while the \gls{dvs} creates a continuous data stream.}
    \label{fig:synchronization}
    \vspace{-1em}
\end{figure}

%
\section{Method}
\label{sec:method}
The proposed system undergoes a comprehensive \textbf{quantitative analysis}, taking into account factors such as sensor shutters, optical distortions, and influences from camera peripherals. Understanding these technical challenges is crucial for effectively \textbf{synchronizing and calibrating} the pixels of the \gls{dvs} to the RGB data. The evaluation of the synchronization is done by testing four different synchronization methods, visualized in \cref{fig:synchronization}, and comparing the calibration accuracies of the matched camera \glspl{fov} on a pixel level \cref{tab:lens_combinations}. This requires a conversion from the event-based data stream into image frames by accumulating event data over a constant period of time \cite{Rebecq19cvpr, Rebecq19pami}. While the calibration process is carried out using Kalibr \cite{furgale2013unified}, utilizing the protocol suggested in \cite{muglikar2021calibrate}, the paper lacks reproducible error measurements. To address this gap, the standard deviation of the re-projection error in pixels---reported by Kalibr---is included in this report, allowing for more comprehensive comparisons and evaluations of calibration accuracy. To verify the homography projection accuracy, edges in the two \gls{dvs} frame and RGB frame views of distant scenes are detected using a Canny filter, and the matching deviation from the center is computed using \gls{zncc} \cite{2005distefanoZNCCbasedTemplateMatchinga}.
The estimation of \gls{dvs} \textbf{data quantity} is non-trivial for different backgrounds and camera movement speeds and differs significantly from the data amount created by frame-based sensors. In the case of event data, the IMX636 sensor can produce anywhere from nearly \SI{0}{Ev/\s } to \SI{1}{\giga Ev/\s} in highly dynamic scenes \cite{EventbasedVisionSensor}. However, in practice, the USB peripheral is limited to $\sim$\SI{1.6}{\giga\bit/\s} transfer speed, sustaining only about \SI{115}{\mega Ev/\s}. Consequently, this work conducts an analysis of data generation that is dependent on the scene, aiming to offer a reference for future considerations in storage design and bandwidth budget. 
Finally, this work evaluates the perceptible \textbf{object size at different distances} with multiple lenses. This goes hand-in-hand with the selection of suitable lenses for the setup.  While it may appear to involve solely basic optics calculations, the field measurements point out that small objects have a tendency to disappear, even when their theoretical size should be sufficient for detection. This phenomenon can be attributed to the fact that the object at a distance fails to induce sufficient changes in pixel intensity for the \gls{dvs} camera to register a data point. This investigation is done for objects traveling with high-speed (\SI{800}{m/s}) and low to medium-speed (\SIrange[range-phrase=--]{0.5}{30}{m/s}).




%
\section{Experimental Results}
The setup introduced in \cref{sec:setup} has been analyzed to optimize the synchronization of event-based and frame-based sensors, object perception distances, and expected event data rates. For data collection, the camera system has been connected to a laptop via USB 3.0 and USB 3.1 GEN1 for the Prophessee EVK4 and XIMEA MQ022CG-CM, respectively. It was found that without setting an explicit \gls{erc} of \SI{100}{MEv/s}, frames and events could be lost. This was adjusted for certain scenes where bursts could be tolerated. If there were more events present in the scene, the sensor saturated which manifested in line artifacts, visible in \cref{fig:lightning} or in the worst case events being completely dropped.
\begin{table}[b]
    \vspace{-1em}
    \ra{1.1}
    \caption{Lens combinations and corresponding calibration accuracies for synchronization methods (see \cref{fig:synchronization}), represented as the standard deviation in pixels.} 
    \centering
    \resizebox{\linewidth}{!}{%
    \begin{tabular}{@{}rrr|rrr|rrr@{}}\toprule
         \multicolumn{3}{c}{\textbf{Lens Configuration}} & \multicolumn{3}{c}{\textbf{Standard Deviation}} & \multicolumn{3}{c}{\textbf{Cross-correlation deviation}}\\
         \textbf{DVS-L} & \textbf{RGB} & \textbf{DVS-R} & \textbf{DVS-L} & \textbf{RGB} & \textbf{DVS-R} &\textbf{Sync 2} &\textbf{Sync 3} & \textbf{Sync 4} \\
         \midrule
         P100 & X50 & P50 & 1.589& 0.797 & 1.279& 4.1 & 4.9 & 4.7   \\ 
         P75 & X50 & P35 & 1.279 & 0.696& 1.016&  4.4 & 3.4 & 2.3 \\ 
         P50 & X12.5 & P8 & 0.768 & 0.493 & 1.254 & 2.1 & 3.3 & 2.9 \\ 
         P35 & X12.5 & P8 & 0.665 & 0.494 & 0.900  & 2.2 & 3.4  & 1.6\\ 
        \bottomrule
    \end{tabular}
    \label{tab:lens_combinations}
    }
    \vspace{-1em}
\end{table}
\begin{table}[b]
    \ra{1.1}
    \caption{Object resolution using the DVS at different distances using a selection of lenses.} 
    \centering
    \resizebox{\columnwidth}{!}{
    \begin{tabular}{@{}rr|rr|rr|rr@{}}\toprule
         & & \multicolumn{2}{c}{\textbf{8mm lens}} & \multicolumn{2}{c}{\textbf{35mm lens}} & \multicolumn{2}{c}{\textbf{100mm lens}}\\
         \textbf{Object} & \textbf{Distance} & \textbf{Theory} & \textbf{Meas.} & \textbf{Theory} & \textbf{Meas.} & \textbf{Theory} & \textbf{Meas.} \\
         Drone \SI{30}{\cm} &     100m & \SI{5}{px} & \SI{5}{px} & \SI{22}{px} & \SI{20}{px} & \SI{61}{px} & - \\
         Drone \SI{30}{\cm} &     300m & \SI{1.6}{px} & ND & \SI{7}{px} &  \SI{8}{px} & \SI{20}{px} & - \\
         Drone$^\ddagger$ \SI{30}{\cm} & 350m & \SI{1.4}{px} & $\sim$\SI{3}{px}$^\ddagger$ & \SI{6}{px} & $\sim$\SI{7}{px} & \SI{18}{px} & - \\
         Bullet \SI{2}{\cm} &      10m & \SI{3}{px} & \SI{3}{px} & \SI{14}{px} & - & \SI{41}{px} & \SI{38}{px} \\
         Bullet \SI{2}{\cm} &      30m & \SI{1}{px} & ND & \SI{5}{px} & \SI{1}{px} & \SI{14}{px} & \SI{1}{px}  \\
         Bullet \SI{2}{\cm} &     100m & \SI{0.3}{px} & ND & \SI{1.4}{px} & ND& \SI{4}{px} & \SI{1}{px}  \\
         Bullet \SI{2}{\cm} &     400m & \SI{0.08}{px} & ND & \SI{0.35}{px} & ND & \SI{1}{px} & ND \\
        \bottomrule
        \multicolumn{8}{l}{$^\ddagger$ Drone with homogeneous blue-white sky background.} \\
        \multicolumn{8}{l}{ND: in configurations where the object is not detectable.}
    \end{tabular}}
    \label{tab:object_resolution}
    \vspace{-0.5em}
\end{table}

\textbf{\textit{A. Calibration accuracy achieved:}} This paper exploits the approach proposed in \cite{muglikar2021calibrate} for event camera calibration using Kalibr and reports the calibration accuracy as reprojection error standard deviation to be below \SI{1.6}{px}. This is significantly worse than what's possible with Global shutter cameras, which readily feature an error below \SI{0.2}{px}. Method 1 shown in \cref{fig:synchronization} proved to generate poor image quality resulting in calibration failing. Methods 2, 3, and 4 had similar performance. As such, the achieved standard deviation error results using only method 2 are displayed in \cref{tab:lens_combinations}. Notably, the calibration of large focal-length lenses proved to be a challenging task. 

\textbf{\textit{B. Lens vs. object size and distance:}} Investigating the objects traveling with different speeds at different distances, a commercial drone (\SI{30}{\cm}) at \SI{300}{\meter} altitude can be detected on an accumulated \gls{dvs} frame with up to \SI{12}{px} using a \SI{35}{\milli\meter} lens as shown in \cref{tab:object_resolution}. Against a homogeneous background (sky), the contrast is much greater. The drone is even detectible at an altitude of \SI{350}{\meter} using the \SI{8}{\milli\meter} lens despite according to optical calculations only having a size of \SI{1.4}{px}. However, when investigating fast-traveling objects like bullet shots---with a size of \SI{2}{\cm}---the bullet is lost at $<$\SI{80}{\meter} even though it still should be \SI{1}{px} at a distance of \SI{400}{\meter} (using 100mm lens).


\textbf{\textit{C. Data amount vs. moving \gls{dvs} camera:}} As described in the method \cref{sec:method}, the amount of data created is non-trivial to calculate. Therefore baseline measurements for a static camera and dynamic camera (panning at $\leq$ \SI{0.25}{\radian/\second}) are reported, measuring bullet shots, flying drones and vehicles, on a homogeneous (sky) and non-homogeneous background (mountain and grass), see \cref{tab:data_amount}. Notably, the dynamic camera produces orders of magnitude more data. However, homogeneous backgrounds, such as the sky, can significantly reduce the number of context events. The highest data rate without reaching saturation is \SI{180}{\mega\byte/\second}. However, for short bursts, the sensor can reach a peak rate of \SI{400}{\mega\byte/\second}, which is the maximal rate set in the \gls{erc}. For comparison, the Ximea frame camera recording in 10-bit RAW generates \SI{260}{\mega\byte/\s} continuously.

\begin{table}[t]
    \ra{1.1}
    \caption{The average data rate created by the Prophesee EVK4 \gls{dvs} cameras for different Scenery and Camera setup scenarios.}
    \begin{center}
    \resizebox{\columnwidth}{!}{    
    \begin{tabular}{@{}rrrrrrrr@{}}\toprule
        & & & \multicolumn{3}{c}{\textbf{Camera motion}} \\
        & & \textbf{Obj. Size} & \textbf{Static} & \textbf{Dynamic } & \textbf{Dynamic - Sky $^\ddagger$} \\
         \midrule
         \multirow{6}{*}{\rotatebox[origin=c]{90}{\textbf{Scenery}}} & \textbf{Drone Near} & \SI{30}{\cm} &  \SI{20}{\mega\byte/\second}  & \SI{70}{\mega\byte/\second} & \SI{36}{\mega\byte/\second}\\ 
        & \textbf{Drone Far} & \SI{30}{\cm}&  \SI{0.5}{\mega\byte/\second}  & \SI{70}{\mega\byte/\second} & \SI{3}{\mega\byte/\second}\\ 
        & \textbf{Car} & \SI{3}{\m}&  \SI{15}{\mega\byte/\second}  & \SI{100}{\mega\byte/\second} & - \\ 
        & \textbf{Shots} & \SI{2}{\cm}&  \SI{0.1}{\mega\byte/\second}  & \SI{60}{\mega\byte/\second} & - \\ 
        & \textbf{Explosion} & \SI{5}{\m}& $^*$\SI{450}{\mega\byte/\second} & - & - \\ 
        & \textbf{Rain} & - &  \SI{160}{\mega\byte/\second}  & $^*$\SI{360}{\mega\byte/\second} & \SI{10}{\mega\byte/\second}\\ 
        \bottomrule
        & &   \multicolumn{3}{l}{\footnotesize{$^*$ Peak value before saturation reached}} \\
     & &   \multicolumn{3}{l}{\footnotesize{$^\ddagger$ Object with homogeneous sky background}}
    \end{tabular}}
    \label{tab:data_amount}
    \end{center}
    \vspace{-2em}
\end{table}

\textbf{\textit{D. System limitation:}}
Last but not least, some limitations of the \gls{dvs} cameras are presented. Most notably, insects and non-homogeneous background movements (e.g. wind moving trees) can create a significant number of irrelevant events. Scenes with brief abrupt changes in light intensity, such as explosions, lightning, strobe lights, or motion can temporarily overwhelm the sensor; sometimes resulting in complete data loss. In particular, the rain proved a significant challenge often causing camera saturation. 


\section{Conclusion}
This work presented a multi-modal camera setup for semi-automated collection and labeling of \gls{dvs} data. This included the analysis of various event-to-RGB image frame synchronization techniques, and corresponding calibration accuracy and pixel cross-correlation are reported. Measurement results show that lenses with similar \gls{fov} and low distortion achieve a calibration accuracy of less than \SI{0.90}{px}, with a corresponding pixel cross-correlation of \SI{1.6}{px}. Furthermore, tests demonstrated that an object of a size of \SI{30}{\cm} on the homogeneous background can still be detected at a distance of \SI{350}{\m}, using the Prophesee EVK4 sensor with a wide \SI{8}{\mm} lens. In-field experience of the data bandwidth ranges from \SI{0.1}{\mega\byte/\second} - \SI{20}{\mega\byte/\second} for static camera recordings while a moving \gls{dvs} camera generates data in the order of \SI{350}{\mega\byte/\second}.

\section*{Acknowledgments}
The authors would like to thank armasuisse Science \& Technology and RUAG Ltd. for funding this research.

\bibliographystyle{IEEEtranDOI}
\bibliography{IEEEabrv,sentinel_references.bib}

\end{document}